\documentstyle[11pt,aaspp4]{article}

\begin{document}

\title{Spallation of Iron in Black Hole Accretion Flows}

\author{Jeffrey G. Skibo}

\affil{E. O. Hulburt Center for Space Research, Code 7653, \\
Naval Research Laboratory, Washington, DC 20375-5352 \\
skibo@osse.nrl.navy.mil}
 
\begin{abstract}

In the local Galactic interstellar medium there is approximate energy
equipartition between cosmic rays, magnetic fields and radiation. If 
this holds in the central regions of AGN, in particular Seyfert
galaxies, then consideral nuclear spallation of Fe occurs, resulting in
enhanced abundances of the sub-Fe elements Ti, V, Cr and Mn. These elements
produce a cluster of X-ray flourescence lines at energies just below the 6.4 
keV Fe-K$\alpha$ line. It is suggested that the red wings on the Fe
lines observed with ASCA from various Seyfert AGN are due to the unresolved
line emission from these elements. Future observations with more sensitive
X-ray instruments should resolve these lines. The estimated gamma ray emission
from nuclear deexcitation and neutral pion production is calculated and found to
be below the sensitivities of any current instruments. However, very luminous
nearby Seyferts displaying Fe lines with red wings would have $>100$ MeV 
continuum emission detectable by future instruments such as
GLAST. 

\end{abstract}


\section{Introduction}

Recent X-ray observations with ASCA of various Seyfert AGN have revealed the
presence of asymmetries in the 6.4 keV Fe-K$\alpha$ line profile (Tanaka et al.
1995; Yaqoob et al. 1995; Mushotzky et al. 1995; Nandra et al. 1996; Iwasawa et
al. 1996). The spectra generally display a narrow 6.4 keV line with a red wing
extending below 5 keV and a sharp cut-off above 6.5 keV. This has been
interpreted as emission from the inner regions of an accretion disk of a
massive black hole, where the asymmetric line profile is attributed to Doppler
and gravitational effects (Tanaka et al. 1995; Fabian et al. 1995). 

Other mechanisms for producing these skewed line profiles have difficulties
(see Fabian et al. 1995 for a detailed discussion). For example, Comptonization
is capable of producing red wings on lines, however the scattering depth
required by the ASCA data is at least 3. This would imprint deep absorption
edges in the spectrum unless the medium was nearly totally ionized. This
requires the medium to be close to the central black hole, where special and
general relativistic effects would dominate anyway. Another possibility is that
the line is intrinsically broad and just the blue wing is absorbed. The problem
is that the Fe absorption edge occurs at energies greater than 7.1 keV, whereas
the ASCA data show a drop off above about 6.5 keV. Other relativistic motions
not associated with accretion disks seem contrived simply because no emission
to the blue is observed. However, it is also puzzling as to why in the
accretion disk scenario the inferred disk orientations are always nearly face
on with respect to the line-of-sight, especially in the case of NGC 4151 where
the host galaxy is viewed nearly edge on (Yaqoob et al. 1995). 

In this paper another scenario for producing the Fe-line red wings is
introduced. Enhanced abundances of the sub-Fe metals titanium, vandium,
chromium and manganese will produce flourescence lines at energies between 4.5
and 6.4 keV. The unresolved detection of these lines would appear in the
spectrum as a red wing on the 6.4 keV line of Fe. The enhancement of sub-Fe
elements occurs naturally through cosmic ray spallation of Fe, the most
abundant metal in astrophysical plasmas. Bombardment of an Fe nucleus by a
proton of energy $\gtrsim10$ MeV tends to chip off a light nucleus, thus
converting the Fe into some isotope of an element just below Fe. Interactions
of this sort will modify the abundances significantly from solar composition.
This effect is observed in the relative elemental abundances of Galactic
cosmic-rays, which are subject to bombardment by boosted ambient interstellar
protons (Lund 1989). Lithium is also abundantly produced by spallation
reactions, and excess Li has been observed in Galactic black hole systems
(Mart\'in et al. 1992). 

In the local interstellar medium there is approximate energy equipartion
between cosmic ray protons, ambient magnetic fields and ambient radiation
fields, the energy densities are all roughly 1 eV cm$^{-3}$. The cosmic-ray gas
and ambient magnetic field are probably in approximate pressure equilibrium,
but there is no known reason for the equipartition of these constituents with
ambient radiation fields, it could just be coincidental. It is shown that if
this situation holds in the nuclear regions of Seyfert AGN then sufficient
spallation of Fe occurs resulting in the appropriate abundances of Ti, V, Cr
and Mn for the requred 4.5-6.4 keV X-ray line emission. 

In \S 2 nuclear spallation rates and abundances enhancements are calculated for
various proton spectra. The X-ray emission resulting from the irradiation of
this material by an incident power law continuum is calculated using a Monte
Carlo simulation in \S 3. Gamma ray emission due to nuclear excition and
neutral pion production is estimated in \S 4. Finally, a brief statement of the
conclusions reached here is given in \S 5. 

\section{Nuclear Spallation Reactions}

It is assumed that the material accreting onto the central black hole enters
the accretion flow with solar composition (Anders \& Grevesse 1989) and that no
further nucleosynthesis via fusion reactions occurs. It is also assumed that
the central source produces energetic protons with luminosity approximately
equal to that of the emergent radiation. This implies energy equipartition
between radiation and energetic protons near the central source. No detailed
assumptions concerning the geometry of the region are made. It is simply
assumed that the protons are homogeniously injected into the accretion flow
which provides a thick target for the protons. This is justified if the
magnetic energy density is comparable that of the protons which then do not
stream freely but rather have highly convoluted trajectories. This
approximation is often employed in Galactic cosmic ray propagation studies and
has been shown to provide an adequate explanation of the isotopic cosmic ray
abundunces (Shapiro \& Silberberg 1974; Engelmann et al. 1990). In this
approximation the steady state proton intensity $J_p(E)$ (protons s$^{-1}$
cm$^{-2}$ ster$^{-1}$ MeV$^{-1}$) in any local region comoving with the flow
satisfies 
\begin{equation}
\frac{d}{dE}\left(\left|\frac{dE}{dx}(E)\right|\,J_p(E)\right)+
\frac{J_p(E)}{\Lambda(E)}
=\frac{1}{4\pi M}\frac{dN_p}{dt\,dE}(E).
\end{equation}
The first term on the left represents continuous energy losses due to Coulomb 
collisions in the ambient medium, which for the purposes here is assumed to 
consist of ionized hydrogen and helium with $n_{He}/n_H=0.1$.
The loss rate is approximated from the curves given in Forman, Ramaty \& 
Zweibel (1986):
\begin{equation}
\left|\frac{dE}{dx}(E)\right|=110 E^{-0.455}\beta^{-1}\qquad\qquad
{\rm (MeV\;gram^{-1}\;cm^2)}.
\end{equation}
The second term represents discrete energy losses due to innelastic nuclear 
collisions. The path length in units of grams cm$^{-2}$ is given by
\begin{equation}
\Lambda(E)=\frac{14m_p}{10\sigma_{pp}(E)+\sigma_{p\alpha}(E)},
\end{equation}
where $\sigma_{pp}(E)$ and $\sigma_{p\alpha}(E)$ are the inelastic $pp$ and
$p\alpha$ cross sections (Meyer 1972). The term on the right is the
rate per energy per accreted mass that protons are injected into the system. 
Here $M$ is the total mass of accretion material, not the central object.
The formal solution of equation (1) is
\begin{equation}
J_p(E)=\frac{1}{4\pi M \left|\frac{dE}{dx}(E)\right|}\int_E^\infty dE'
\frac{dN_p}{dt\,dE}(E')\,\exp\left(-\int_E^{E'}dE''
\frac{1}{\left|\frac{dE}{dx}(E'')\right|\Lambda(E'')}\right)
\end{equation}

It is assumed that the injection spectrum is a simple power law in 
kinetic energy,
\begin{equation}
\frac{dN_p}{dt\,dE}(E)=(s-2)\eta\dot{M}c^2E_0^{s-2}E^{-s},
\end{equation} 
where the spectral index satisfies $s>2$. This is in accord with the source
spectrum of Galactic cosmic rays inferred from measurements of the cosmic ray
spectrum at the top of the atmosphere (Berezinski et al. 1990). Because the
spectrum is divergent at lower energies it is necessary to cut the spectrum off
below some energy $E_0$. The normalization in equation (5) assures a total
proton luminosity $L_p=\eta\dot{M}c^2$ where $\eta$ represents the efficiency
for converting the rest mass of the accretion flow into energetic particles.
The proton luminosity is assumed to be approximately equal to the radiation
luminosity where $\eta$ is generally assumed to be on the order of 0.1. 

The mean rate per target nucleus for spallation of species $i$ into species 
$j$ by energetic protons is given by
\begin{equation}
R_{i\rightarrow j}=4\pi\int_0^\infty dE\;\sigma_{i\rightarrow j}(E)\,J_p(E),
\end{equation} 
where $\sigma_{i\rightarrow j}$ is the appropriate spallation cross section.
In this work the empirical formulas of Letaw, Silberberg, \& Tsao (1983)
are used for all spallation cross sections. Combining equations (4), (5) and 
(6) yields
\begin{eqnarray}
\frac{M}{\dot{M}}R_{i\rightarrow j}&=&(s-2)\eta c^2E_0^{s-2}
\int_0^\infty dE\;\sigma_{i\rightarrow j}(E) 
\cr
&&\quad\cdot\frac{1}{\left|\frac{dE}{dx}(E)\right|}\int_E^\infty dE'
E'^{-s}\exp\left(-\int_E^{E'}dE''
\frac{1}{\left|\frac{dE}{dx}(E'')\right|\Lambda(E'')}\right).
\end{eqnarray}
The quantity on the left hand side is the spallation rate per target nucleus
multiplied by the accretion time scale $t_a\equiv M/\dot{M}$. This is a 
useful quantity for what follows. The important thing is that it only depends 
on the input proton spectrum and the nuclear physics of spallation, and not
on the details of the accretion flow. 

Spallation of solar abundance material will generally deplete iron, which is
the most abundant metal, and enhance the abundances of sub-Fe metals.  The
focus in this work is on those spallation products which produce flourescence
lines in the energy range 4.5-6.4 keV. The K$\alpha$ lines of titanium,
vanadium, chromium and manganese are at energies 4.5, 4.9, 5.4 and 5.9 keV,
respectively. The reaction network considerd here will consist of all stable
isotopes of Ti, V, Cr, Mn and Fe plus any unstable isotopes which
decay into these (see Table 1). 

The system of continuity equations regulating the abundances of 
Fe and its spallation products are
\begin{mathletters}
\begin{eqnarray}
\frac{\partial N_{Fe}}{\partial t}&=&-N_{Fe}R_{Fe}
\\
\frac{\partial N_i}{\partial t}&=&-N_iR_i+N_{Fe}
\left(R_{Fe\rightarrow i}+\sum_j R_{Fe\rightarrow j(i)}\right)
\end{eqnarray}
\end{mathletters}
where the index $i$ extends over all stable isotopes of Ti, V, Cr and Mn given
in the second column of Table 1. The indicies $j(i)$ in the summation extend
over all unstable isotopes which decay into the stable isotope $i$. These are
given in the first column of Table 1. The solutions of equations (8a) and (8b)
are 
\begin{mathletters}
\begin{eqnarray}
N_{Fe}(t)&=&N_{Fe}^{\odot}e^{-R_{Fe}t}
\\
N_i(t)&=&N_i^{\odot}e^{-R_it}+N_{Fe}^{\odot}
\left(e^{-R_{Fe}t}-e^{-R_it}\right)
\left(\frac{R_{Fe\rightarrow i}+\sum_j R_{Fe\rightarrow j(i)}}
{R_i-R_{Fe}}\right).
\end{eqnarray}
\end{mathletters}
The elemental abundances relative to solar values are obtained by summing 
over the different isotopes of each element,
\begin{equation}
X_Z(t)=\frac{1}{N_Z^\odot}\sum_{\{A\}_Z} N_{A,Z}(t)
\end{equation}
where $\{A\}_Z$ means the sum extends over all the stable isotope listed in 
Table 1 belonging to element $Z$. 
The initial relative elemental abundances are taken from Anders \& Grevesse 
(1989) and the initial abundnaces of each isotope in a given element are 
taken from Lederer \& Shirley (1978) also given
for the relevant stable end products in Table 1. 

The procedure is as follows: The time scale is set equal to the accretion time
scale $t_a=M/\dot{M}$; Equation (7) is used for the product of the spallation
rates with this time scale for a given proton injection spectrum paramaterized
by $s$, $E_0$ and $\eta$; Equations (9a), (9b) and (10) are then used to obtain
the elemental abundances relative to solar. 

In Figure 1 relative abundances are shown as a function of the proton spectral
index $s$ for $E_0=10$ MeV and $\eta=0.1$. It can be clearly seen that the
sub-Fe elements are enhanced at the expense of Fe. This is most pronounced for
proton spectral indices in the range 2-3. The relative abundances depend on the
assumed low energy cut-off $E_0$ as shown in Figure 2 where $\eta=0.1$ and
$s=2.3$. Between 10-100 MeV the dependence is rather weak until about 1
Gev where there is generally less spallation. The dependence on the efficiency
factor $\eta$ is shown in Figure 3 where $s=2.3$ and $ E_0=10$ MeV. The
production of Ti, V, Cr and Mn peak at efficiencies on the order of 10\%. 

In summary, the abundance enhancements over solar of Ti, V, Cr and Mn due to
the spallation of Fe by energetic protons with power law spectra are greatest
for proton spectral indicies in the range 2.0-2.5, low energy cut-offs between
10 MeV - 1 GeV and efficiencies on the order of 10\%. In the following the 
abundance enhancements resulting from a proton spectrum with $s=2.3$, 
$E_0=10$ MeV and $\eta=0.1$ are used to calculate the X-ray line emission.

\section{X-ray Emission}

A Monte Carlo simulation is used to calculate the X-ray line emission. It is
assumed that a central point source isotropically emits a simple power law
X-ray continuum with energy spectral index $-1$ (photon spectral index $-2$).
This is in accord with observations which show that the average intrinsic X-ray
power law index of Seyferts is found to be close to $-0.9$ when the spectral
hardening above 10 keV due to reflection by cold optically thick material in
the vicinity of the source is taken into account (Nandra \& Pounds 1994).
However, the spectral hardening $\gtrsim10$ keV attributed to reflection could
be an intrinsic feature of the central source (Skibo \& Dermer 1995). Any
hardening already present in the central source would just add slightly to the
line emission, so for simplicity a simple power law with energy index $-1$ is
used. Material with enhanced abundances of sub-Fe elements as calculated in the
last section is assumed to surround the central source.  For simplicity, the
material is assumed to have spherical geometry and uniform density. In addition
to the abundances, the sphere of material is characterized by the Thomson
optical depth from the center to the surface. 

In the Monte Carlo simulation photons are injected at the center of the sphere
with energy selected from the initial power law distribution. In the energy
range considered here (3-30 keV) the relevant processes are photoelectric
absorption, flourescence line emission  and Compton scattering. A photon is
followed until it either scatters or is absorbed where the probabilities for
these events are derived from the total cross sections. For Compton scattering
the full Compton cross section is used. For photoelectric absorption the
subroutines of Baluci\'nska-Church \& McCammon (1992) are supplemented with the
cross sections for the elements Ti, V and Mn taken from Henke et al. (1982).
The abundances are assumed to be solar (Anders and Grevesse 1989) except for
the modifications of Ti, V, Cr, Mn and Fe derived in the last section. If a
scattering event occurs, the photon is reinjected at the scattering location
with a new direction and energy drawn from probability distributions derived
from the differential Compton cross section. If the photon is absorbed the
element responsible for the absorption is determined from the absorption cross
sections. If the absorbing element is Ti, V, Cr, Mn or Fe then the K$\alpha$
flourescence yields ($Y_{Ti}=0.219$, $Y_{V}=0.250$, $Y_{Cr}=0.282$,
$Y_{Mn}=0.314$ and $Y_{Fe}=0.347$; Lederer \& Shirley 1978) are used to
determine whether or not a K$\alpha$ line photon is emitted. If a K$\alpha$
line photon is emitted then it is reinjected at the location of the absorbing
event with the relevant K$\alpha$ line energy and randomly chosen direction.
Ultimately, a photon escapes the system in which case it is recorded and
binned, or it is absorbed without the emission of a flourescence line photon.
The process is repeated until a statistically good spectrum is obtained. 

In Figure 4 the emergent X-ray spectrum is shown for a reprocessing sphere of
Thomson radial depth 0.1. The Ti, V, Cr, Mn and Fe abundances are those
resulting from a proton spectrum with $s=2.3$, $E_0=10$ MeV and $\eta=0.1$. The
6.4 keV Fe line is still prominant, but now with the spallation enhanced
abundances, the Cr and Mn lines at 5.4 keV and 5.9 keV, respectively, are also
clearly above the continuum. There is also a hint of the 4.5 keV (Ti) and 4.9
keV (V) lines. The line fluxes relative to Fe are 4\% (Ti), 3\% (V), 33\% (Cr)
and 11\% (Mn) which sum to 51\% of the Fe-line flux. This spectrum should not 
be regarded as a detailed model for the X-ray emission from Seyfert AGN. It is
displayed simply to demonstrate that the lines from the sub-Fe elements will
protrude above the continuum with combined strength comparable to that of the
Fe-line. 

Nandra et al. (1996) have produced a continuum subtracted mean line profile
from the ASCA observations of a sample of 18 Seyfert 1 AGN. These data are
shown in Figure 5 along with a simple model consisting of the sum of 5
gaussians each of width 0.4 keV centered at the k$\alpha$ energies of the
elements Ti, V, Cr, Mn and Fe. In this fit the fluxes in the gaussian lines of
the sub-Fe elements relative to that of Fe are found to be 4\% (Ti), 10\% (V),
33\% (Cr) and 25\% (Mn). Thus the flux in the sub-Fe elements adds up to about
72\% of that in the Fe line. The data seem to require more V and Mn than that
produced in the simple model presented here but otherwise are in qualitative
aggreement with it. A more detailed analysis should include Ni and the
secondary spallation of Cr which would result in more V and Mn. But another
possibility could be that the lines from the spalled material are intrinsically
broader and that there is an additional component of narrow Fe-line emission.
The important point is that, when corrections are made for the instrumental
response shown in Fig 5 by the dotted curve centered at 6.4 keV, the intrinsic
line widths are 0.35 keV, corresponding to velocities near 20,000 km/s. For
Keplerian motion these velocities are obtained at approximately 300
Schwarzschild radii. Hence, this scenario still requires a compact nucleus, but
the constraints on the size are relaxed by over an order of magnitude from
those imposed by the disk-line model, which requires the emission region to be
within 10 Schwarzschild radii (Tanaka et al. 1995; Fabian et al. 1995). 

\section{Gamma Ray Emission}

Interactions of energetic protons will produce gamma rays. It is well known
from observations of our own Galaxy that cosmic ray protons result in gamma ray
continuum emission above 100 MeV (Bloemen 1989). Furthermore, nuclear
excitation of ambient material by cosmic rays also results in nuclear line
emission in the MeV range (Ramaty, Kozlovsky \& Lingenfelter 1996). Such line
emission has been observed from Solar flares (Share \& Murphy 1995) and the
Orion molecular cloud complex (Bloemen et al. 1994; Ramaty, Kozlovsky \&
Lingenfelter 1995; 1996). 

The continuum emission above 100 MeV is produced mainly by strong
interactions of protons with ambient nuclei resulting in neutral pions
which subsequently decay into photons. The gamma ray production is calculated
using the cross sections and codes developed by Dermer (1986).
Using the form for the proton spectrum given in equation (5) 
the gamma ray emission is given by
\begin{eqnarray}
\frac{dN_\gamma}{dt\,d\epsilon}(\epsilon)&=&
\dot{N}(s-2)\eta c^2E_0^{s-2}
\int_0^\infty d\epsilon\;
\frac{d\sigma_{pH\rightarrow \gamma's}}{d\epsilon}(\epsilon,E)
\cr
&&\qquad\cdot\frac{1}{\left|\frac{dE}{dx}(E)\right|}\int_E^\infty dE'
E'^{-s}\exp\left(-\int_E^{E'}dE''
\frac{1}{\left|\frac{dE}{dx}(E'')\right|\Lambda(E'')}\right).
\end{eqnarray}
where $\dot{N}=\dot{M}/m_p$ is the baryon accretion rate. Using parameters
typical of Seyferts and the proton spectrum with $s=2.3$, $E_0=10$ MeV and
$\eta=0.1$, the integrated $>100$ MeV gamma ray flux at earth is estimated 
to be
\begin{equation}
{\rm Flux}(>100\;{\rm MeV})\simeq 
10^{-7} \left(\frac{L}{10^{44}\;{\rm erg}}\right) 
\left(\frac{d}{1\;{\rm Mpc}}\right)^{-2} 
\qquad {\rm (photons\;s^{-1})}.
\end{equation}
This falls below the $2\sigma$ flux upper limits of Seyferts obtained with
EGRET (Lin et al. 1993), but not by much. A very luminous ($\gtrsim10^{44}$)
nearby Seyfert displaying the skewed Fe line profile could be observable at
high energies with the proposed GLAST instrument if indeed the red wing on the
Fe line is due to the unresolved line emission from sub-Fe elements. 

The nuclear deexcitation line emission is estimated from the yields per proton
calculated for solar flares (Murphy 1985). The emission rate of narrow nuclear
line photons is approximately 
\begin{equation}
\frac{dN_\gamma}{dt}\simeq 
10^{-3}\int_{30\;{\rm MeV}}^\infty dE\;\frac{dN_p}{dt\,dE}(E)
\end{equation}
For the proton spectrum assumed in this analysis the line flux at earth is 
estimated to be
\begin{equation}
{\rm Flux(narrow\;lines)}\simeq 
10^{-6} \left(\frac{L}{10^{44}\;{\rm erg}}\right) 
\left(\frac{d}{1\;{\rm Mpc}}\right)^{-2} 
\qquad {\rm (photons\;s^{-1})}.
\end{equation}
Again, this is well below the sensitivities of any instruments ever built.
This also falls below the sensitivity of the planned low energy gamma ray 
spectrometer INTEGRAL.

\section{Conclusion}

The assumption of approximate energy equipartition between energetic protons,
magnetic fields and radiation in the central regions of Seyfert AGN
implies that consideral nuclear spallation of Fe must occur in the accretion
flow. This results in enhanced abundances relative to solar of the sub-Fe
elements Ti, V, Cr and Mn. The K$\alpha$ line emission of these elements
provides an explanation for the asymmetries on the 6.4 keV Fe-K$\alpha$ lines
observed from various Seyferts with ASCA. The energetic protons also produce
gamma ray line and continuum emission from nuclear deexcitation and neutral
pion production. This emission is estimated to be below the sensitivities of
current gamma ray instruments, however, very luminous nearby Seyferts
displaying Fe-line red wings could be observed by future instruments such as
GLAST.

\acknowledgments 
I thank C. H. Tsao for providing subroutines for the spallation cross sections,
Chuck Dermer for his pion code, Bernard Phlips, Ron Murphy, Brad Graham, Greg 
Jung and Tahir Yaqoob for useful discussions.

\begin{deluxetable}{ll}
\tablenum{1}
\tablewidth{0pt}
\tablecaption{Spallation Products of Fe}
\tablehead{
\colhead{Unstable Products}   & \colhead{Stable End Products}\nl
\colhead{}   & \colhead{(Natural Isotopic Abundance Percentages)}}
\startdata
$^{46}$V, $^{46}$Cr &
~~~~~~~~~~~~~~~~~~~~~$^{46}$Ti (8.2) \nl
$^{47}$V, $^{47}$Cr &
~~~~~~~~~~~~~~~~~~~~~$^{47}$Ti (7.4)\nl
$^{48}$V, $^{48}$Cr &
~~~~~~~~~~~~~~~~~~~~~$^{48}$Ti (73.7)\nl
$^{49}$V, $^{49}$Cr &
~~~~~~~~~~~~~~~~~~~~~$^{49}$Ti (5.4)\nl
$\cdots$&
~~~~~~~~~~~~~~~~~~~~~$^{50}$Ti (5.2)\nl
$\cdots$&
~~~~~~~~~~~~~~~~~~~~~$^{50}$V (0.25)\nl
$^{50}$Mn &
~~~~~~~~~~~~~~~~~~~~~$^{50}$Cr (4.35) \nl
$^{51}$Ti, $^{51}$Cr &
~~~~~~~~~~~~~~~~~~~~~$^{51}$V (99.75) \nl
$^{52}$Mn &
~~~~~~~~~~~~~~~~~~~~~$^{52}$Cr (83.8)\nl
$^{53}$Mn &
~~~~~~~~~~~~~~~~~~~~~$^{53}$Cr (9.5)\nl
$^{54}$Mn &
~~~~~~~~~~~~~~~~~~~~~$^{54}$Cr (2.36)\nl
$^{55}$Fe &
~~~~~~~~~~~~~~~~~~~~~$^{55}$Mn (100)\nl
\enddata
\end{deluxetable}

\clearpage

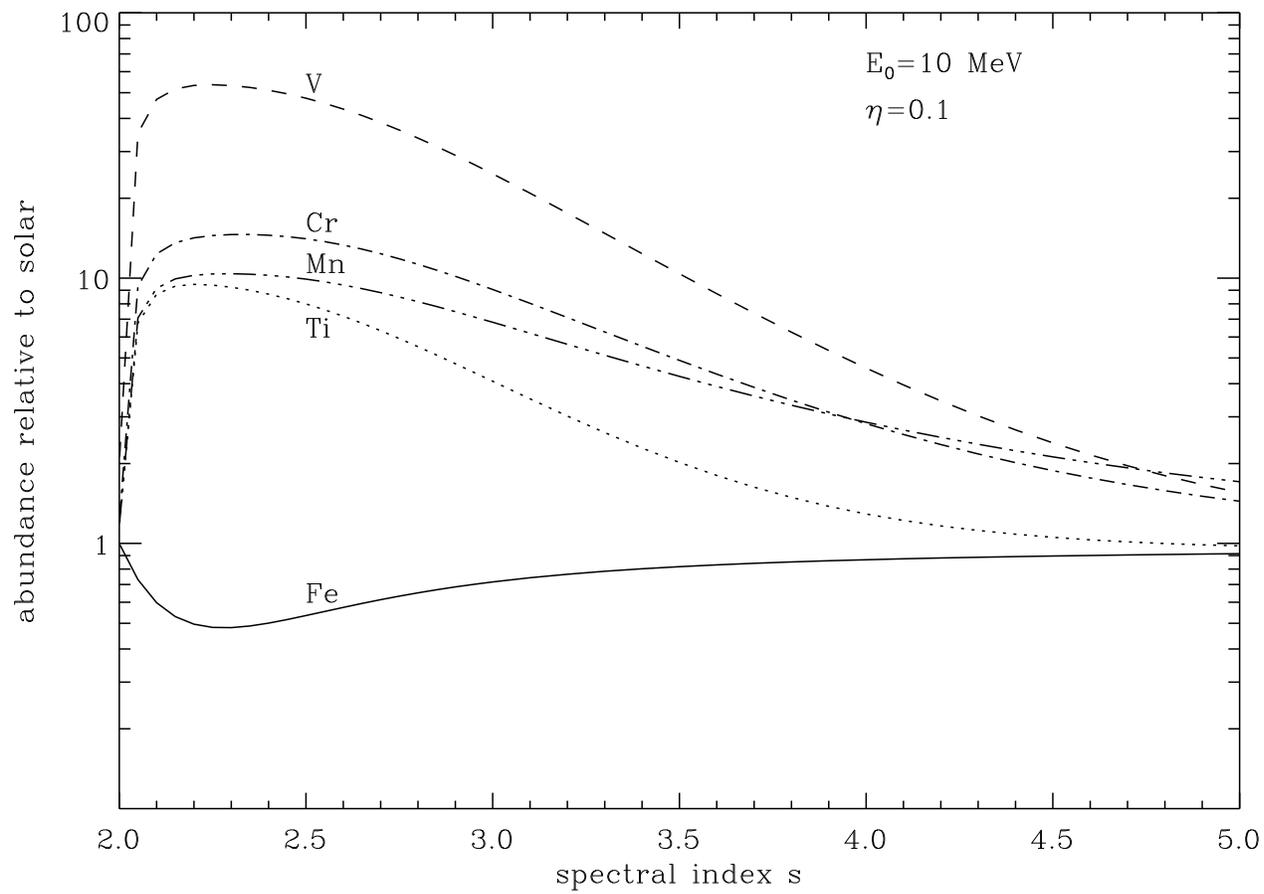
\noindent Fig. 1 --- The abundances of Ti, V, Cr, Mn and Fe relative to their
values in the solar photosphere (Anders \& Grevesse 1989) versus the spectral
index of the power law proton injection spectrum. The proton spectrum is cut
off below $E_0=10$ MeV and the efficiency of converting the accretion mass into
energetic protons is $\eta=0.1$. 

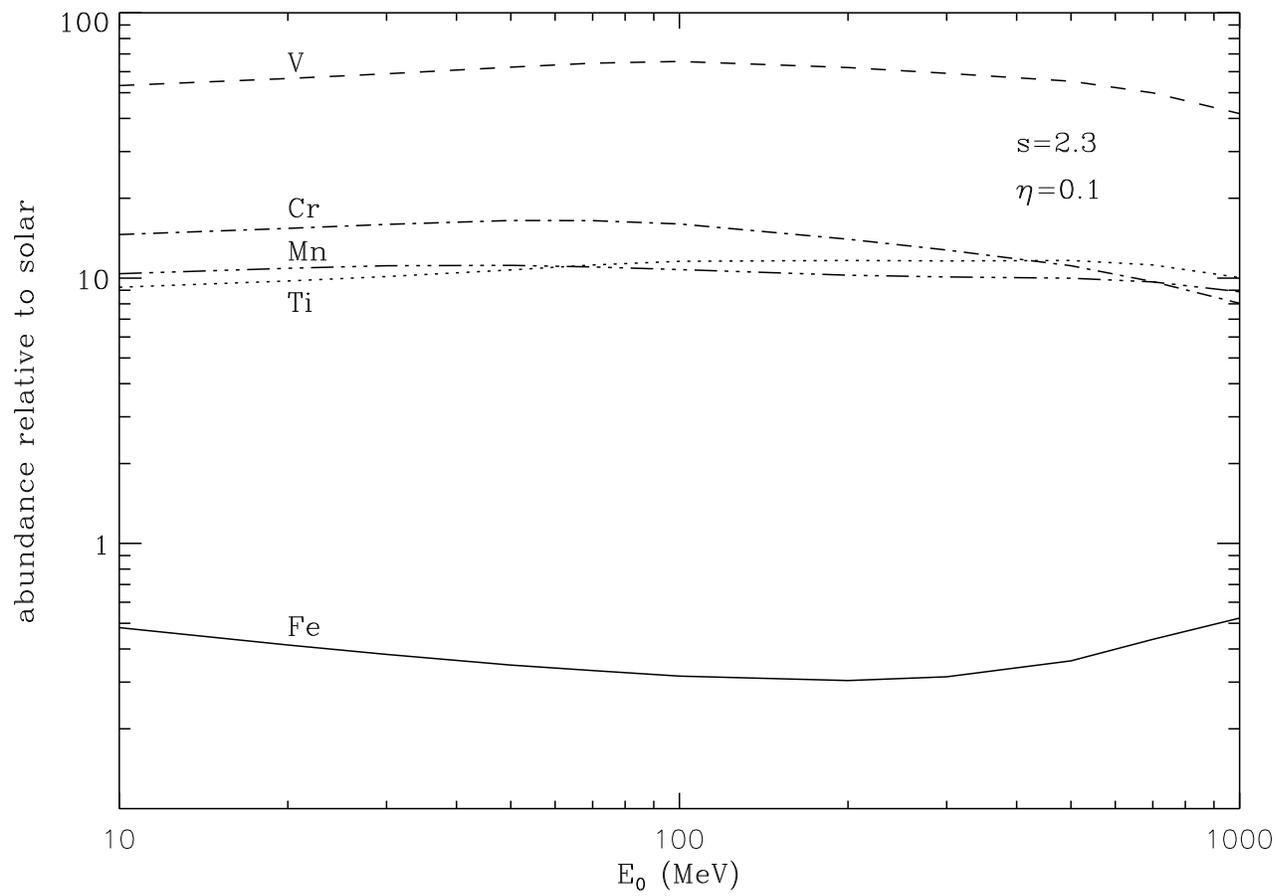
\noindent Fig. 2 --- The abundances of Ti, V, Cr, Mn and Fe relative to their
values in the solar photosphere (Anders \& Grevesse 1989) versus the low energy
cut off of the power law proton injection spectrum. The proton spectral index
is $s=2.3$ and the efficiency of converting the accretion mass into energetic
protons is $\eta=0.1$.

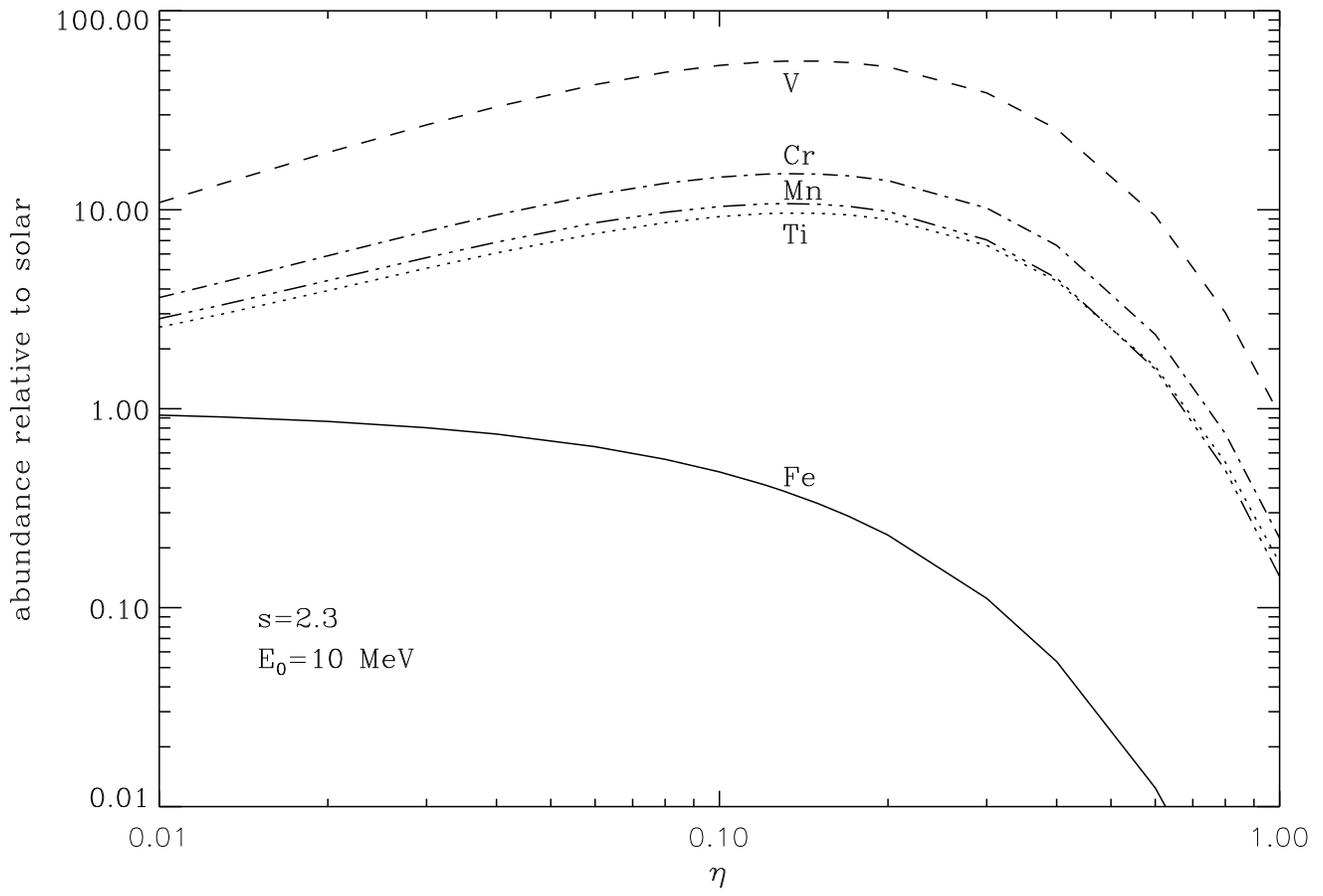
\noindent Fig. 3 --- The abundances of Ti, V, Cr, Mn and Fe relative to their
values in the solar photosphere (Anders \& Grevesse 1989) versus the 
efficiency of converting the accretion mass into energetic
protons. The protons are
cut off below $E_0=10$ MeV and the proton spectral index
is $s=2.3$. 

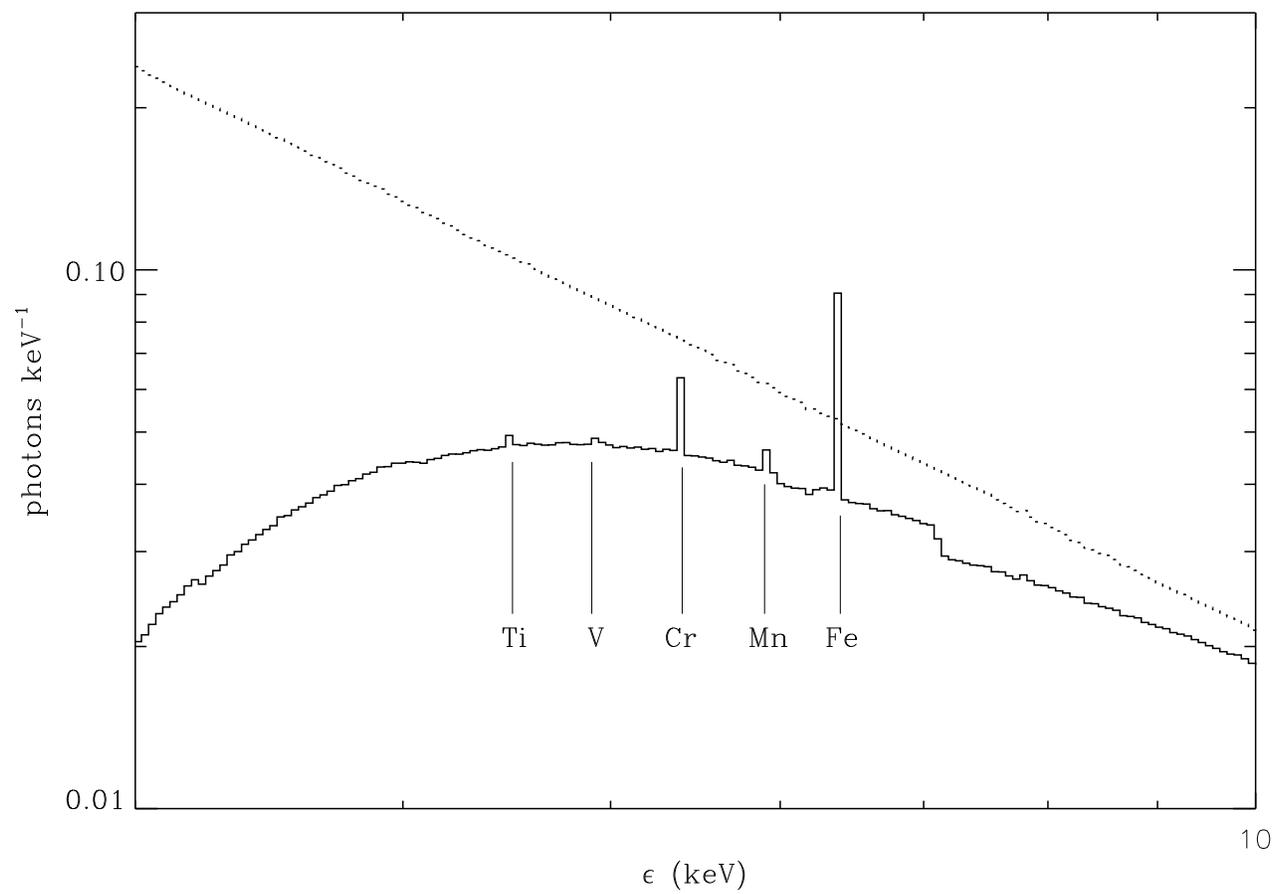
\noindent Fig. 4 --- The X-ray spectrum resulting from a Monte Carlo 
simulation of the emission from a central source with power law 
spectrum (energy index $-1$) passing through a sphere of Thomson scattering 
depth 0.1. The abundances are solar except for Ti, V, Cr, Mn and Fe which are
modified by the linear factors: 9.2 (Ti); 53 (V); 15 (Cr); 10 (Mn); and 
0.48 (Fe). The dotted histogram is the spectrum of the central source 
normalized to 1 photon and the solid histogram is the reprocessed spectrum.

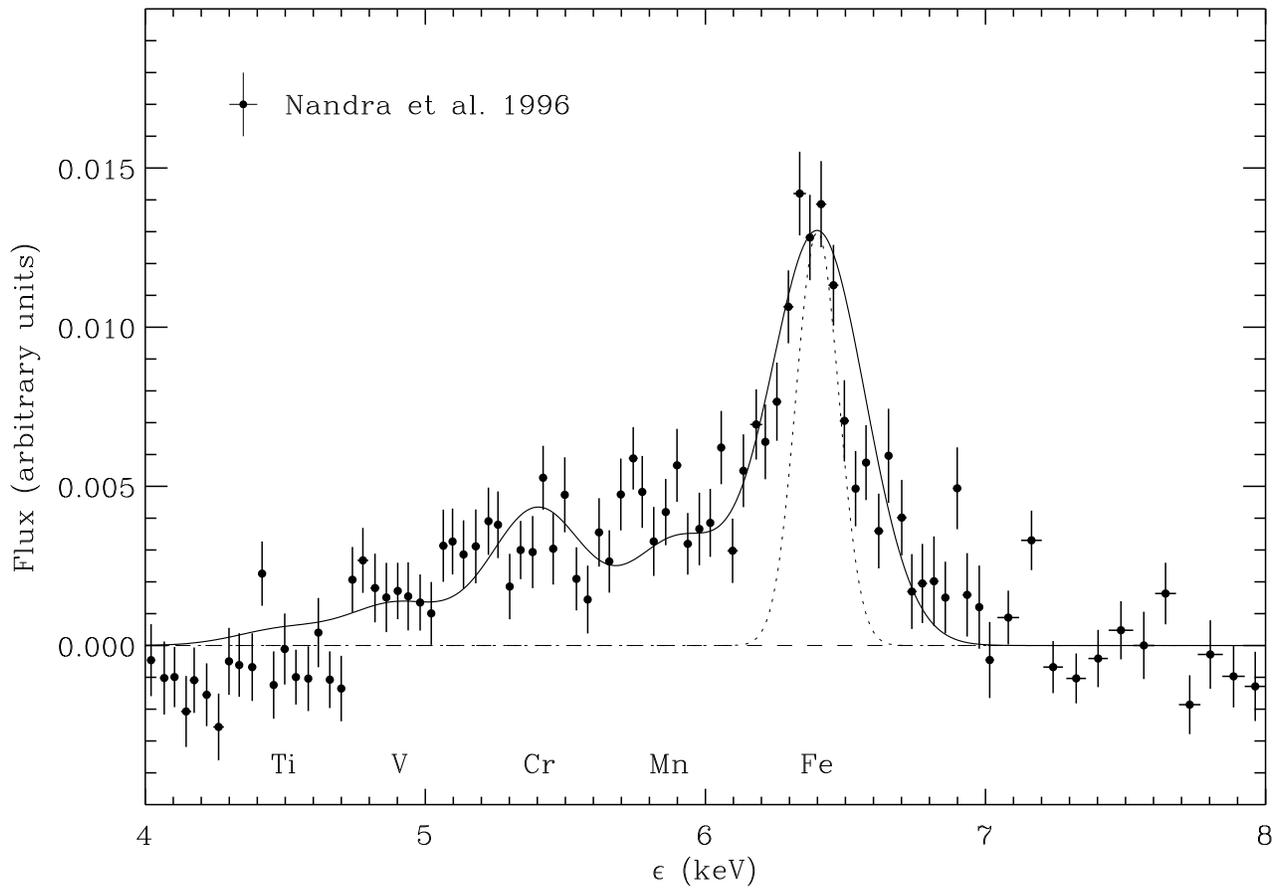
\noindent Fig. 5 --- The data represent the continuum subtracted mean line
profile for a sample of 18 Seyfert 1 AGN observed with the SIS on ASCA (Nandra
et al 1996). The solid curve is the sum of 5 gaussians each of width 0.4 keV
centered at the K$\alpha$ energies of the elements Ti, V, Cr, Mn and Fe. The
fluxes in the gaussian lines of the sub-Fe elements relative to that of Fe are
4\% (Ti), 10\% (V), 33\% (Cr) and 25\% (Mn). The
instrumental response is shown by the dotted curve centered at 6.4 keV. 

\end{document}